\documentclass[11pt, conference]{IEEEtran}
\IEEEoverridecommandlockouts
\usepackage{fancyhdr}
\usepackage{graphicx}
\usepackage{amsmath}
\usepackage{amssymb}
\usepackage{wrapfig}
\usepackage{soul}
\usepackage{xcolor}
\usepackage{verbatim}
\usepackage{subcaption}
\usepackage{tikz}
\usetikzlibrary{spy}
\usepackage{algorithm}
\usepackage{textcomp}
\usepackage[export]{adjustbox}
\usepackage{siunitx}
\usepackage{array}
\usepackage{commath}
\usepackage{url}
\usepackage{paralist}
\usepackage{booktabs} 
\usepackage{algpseudocode}
\usepackage{gensymb}
\usepackage{enumitem}
\usepackage{cancel}
\usepackage{todonotes}
\usepackage{pdfpages}

\setlist{nolistsep}
\graphicspath{ {./figs/} }
\usepackage[switch]{lineno}

\def\BibTeX{{\rm B\kern-.05em{\sc i\kern-.025em b}\kern-.08em
    T\kern-.1667em\lower.7ex\hbox{E}\kern-.125emX}}

\begin{document}



\title{ORBIT: Oak Ridge Base Foundation Model for \\ 
Earth System Predictability \thanks{
This manuscript has been authored by UT-Battelle, LLC, under contract DE-AC05-00OR22725 with the US Department of Energy (DOE). The U.S. government retains and the publisher acknowledges that the US government retains a nonexclusive worldwide license to publish or reproduce the published form of this manuscript, or allow others to do so for US government purposes. DOE will provide public access under the DOE Public Access Plan (http://energy.gov/downloads/doe-public-access-plan).
}
}

\author{
\IEEEauthorblockN{Xiao Wang\IEEEauthorrefmark{1},
Siyan Liu\IEEEauthorrefmark{1},
Aristeidis Tsaris\IEEEauthorrefmark{1},  
Jong-Youl Choi\IEEEauthorrefmark{1},
Ashwin M. Aji\IEEEauthorrefmark{2},
Ming Fan\IEEEauthorrefmark{1}, \\
Wei Zhang\IEEEauthorrefmark{1},
Junqi Yin\IEEEauthorrefmark{1},
Moetasim Ashfaq\IEEEauthorrefmark{1},
Dan Lu\IEEEauthorrefmark{1}, 
Prasanna Balaprakash\IEEEauthorrefmark{1}
}
\IEEEauthorblockA{\IEEEauthorrefmark{1}Oak Ridge National Laboratory, Oak Ridge, United States
\\\{wangx2, lius1, tsarisa, choij, fanm, zhangw3, yinj, mashfaq, lud1, pbalapra\}@ornl.gov}
\IEEEauthorblockA{\IEEEauthorrefmark{2}AMD Research and Advanced Development, Santa Clara, United States
\\ashwin.aji@amd.com}
}

\maketitle

\thispagestyle{fancy}
\lhead{}
\rhead{}
\chead{}
\lfoot{\footnotesize{
SC22, November 13-18, 2022, Dallas, Texas, USA
\newline 978-1-6654-5444-5/22/\$31.00 \copyright 2022 IEEE}}
\rfoot{}
\cfoot{}
\renewcommand{\headrulewidth}{0pt}
\renewcommand{\footrulewidth}{0pt}

\thispagestyle{plain}

\pagestyle{plain}

\maketitle

\begin{abstract} 

Earth system predictability is challenged by the complexity of environmental dynamics and the multitude of variables involved. Current AI foundation models, although advanced by leveraging large and heterogeneous data, are often constrained by their size and data integration, limiting their effectiveness in addressing the full range of Earth system prediction challenges. To overcome these limitations, we introduce the Oak Ridge Base Foundation Model for Earth System Predictability (ORBIT), an advanced vision transformer model that scales up to 113 billion parameters using a novel hybrid tensor-data orthogonal parallelism technique. As the largest model of its kind, ORBIT surpasses the current climate AI foundation model size by a thousandfold. Performance scaling tests conducted on the Frontier supercomputer have demonstrated that ORBIT achieves 684 petaFLOPS to 1.6 exaFLOPS sustained throughput, with scaling efficiency maintained at 41\% to 85\% across 49,152 AMD GPUs. These breakthroughs establish new advances in AI-driven climate modeling and demonstrate promise to significantly improve the Earth system predictability.   

\end{abstract}

\section{Problem Overview}
\label{sec:problem_overview}

Earth system predictability, including tasks such as accurate prediction of extreme weather and climate events, is crucial for safeguarding society and protecting economic stability, as it enables timely preparedness and mitigation strategies. The simulation of such events, however, is challenging as it demands a large ensemble size to accurately represent the diversity of possible scenarios, especially those with a low probability of occurrence. Consequently, there is a pressing need for rapid and accurate Earth system models to ensure that forecasts are timely without compromising prediction accuracy for effective decision-making.

In the Earth system, what unfolds at sub-daily to daily weather scale eventually averages out and can be characterized as sub-seasonal to seasonal climate variations. Gradual shifts in weather and seasonal patterns determine the longer-term climate changes in response to internal and external forcings. Although the Earth system operates seamlessly across different scales, the current physics modeling approach involves using separate prediction systems for each scale, with one system used for weather forecasting, another for sub-seasonal to seasonal prediction, and a third for long-term decadal to centennial climate projections.


In response to the above challenges, AI-based models are emerging as a promising approach and aim to integrate prediction at different scales into a single system. These models leverage advanced machine learning techniques to handle the intricacies of the Earth system  data, potentially revolutionizing our ability to forecast and respond to climate change.
Notable task-specific models like FourCastNet~\cite{fourcastnet}, GraphCast~\cite{graphcast}, Pangu-Weather~\cite{panguweather}, and Stormer~\cite{stormer} primarily use consistent datasets such as ERA5~\cite{era5} for weather forecasting, which limits their flexibility and applicability across varied tasks. Their reliance on homogeneous training and testing data is a significant challenge, restricting their utility for diverse Earth system predictability applications. Moreover, Earth system modeling benefits from abundant simulation data from multi-model resemble projects such as CMIP6~\cite{cmip}, which provide deep insights into the system's structure, function, and dynamics. However, these rich datasets are underutilized in advancing Earth system modeling capabilities. Addressing this gap by integrating such extensive data can markedly improve AI model robustness and applicability, pushing the boundaries of what these tools can achieve in Earth sciences.


Recently, an AI foundation model based on the vision transformer (ViT)   architecture, called ClimaX~\cite{nguyen2023climax}, has been proposed for both weather and climate modeling. It enables pre-training on CMIP6 multi-model simulation data and can be fine-tuned on limited labeled data for a variety of prediction tasks, including weather forecasting, sub-seasonal to seasonal prediction, and multi-decadal climate projection. Despite of ClimaX's promising prediction performance, it has a relatively small model size of 115 million parameters and considers a limited number of variables (also known as channels) from selected atmospheric layers. These limitations, in turn, constrain the ClimaX model's capacity for accurate prediction and scalability potential on large leadership-class HPC systems. 

Scaling ViTs at the pre-training phase is more challenging than natural language processing (NLP) transformer models due to several key factors: 1) ViTs process high-dimensional image data that requires significantly more computational power and memory compared to the one-dimensional text data handled by NLP; 
2) ViTs need to capture complex spatial dependencies within images, which is more computationally intensive than managing sequential dependencies in text; 3) The higher resolution and complexity of image data necessitate greater computational resources and sophisticated processing techniques to scale effectively; 4) Efficiently leveraging hardware for ViTs involves more complex parallelization and optimization strategies due to the intricate nature of image processing; and 5) Existing scaling strategies are mainly optimized for NVIDIA GPUs with fast interconnects, with limited support for AMD or other non-NVIDIA hardware. This results in suboptimal performance when using alternative platforms. Furthermore, optimizations are heavily biased towards NLP transformers, neglecting the specific needs of ViTs on these platforms.

We develop ORBIT, Oak Ridge Base Foundation Model for Earth System Predictability. We designed ORBIT to (1) 
enable scaling the model size up to 113 billion parameters with throughput up to 1.6 exaFLOPS, (2) incorporate 91 channels of climate variables, and (3) pre-train on 10 different CMIP6 datasets~\cite{cmip} with 1.2 million observation data points. In comparison, the current state-of-the-art ClimaX foundation model uses 115 million parameters, 48 variables and 5 CMIP6 datasets. Our reported 113-billion model represents the largest dense ViTs to date, surpassing the existing largest dense vision model~\cite{dehghani2023scaling} by five-fold, and a thousand-fold larger than the largest AI foundation model for climate---ClimaX~\cite{nguyen2023climax} and two hundred times larger than task-specific Stormer model~\cite{stormer}.
ORBIT achieves 1.6 exaFLOPS/684 PFLOPS for the 10/113 billion parameter models on 49,152 Frontier AMD GPUs using mixed-single BFLOAT16 precision.

\begin{figure}[t]
\centering
\includegraphics[width=.8\linewidth]{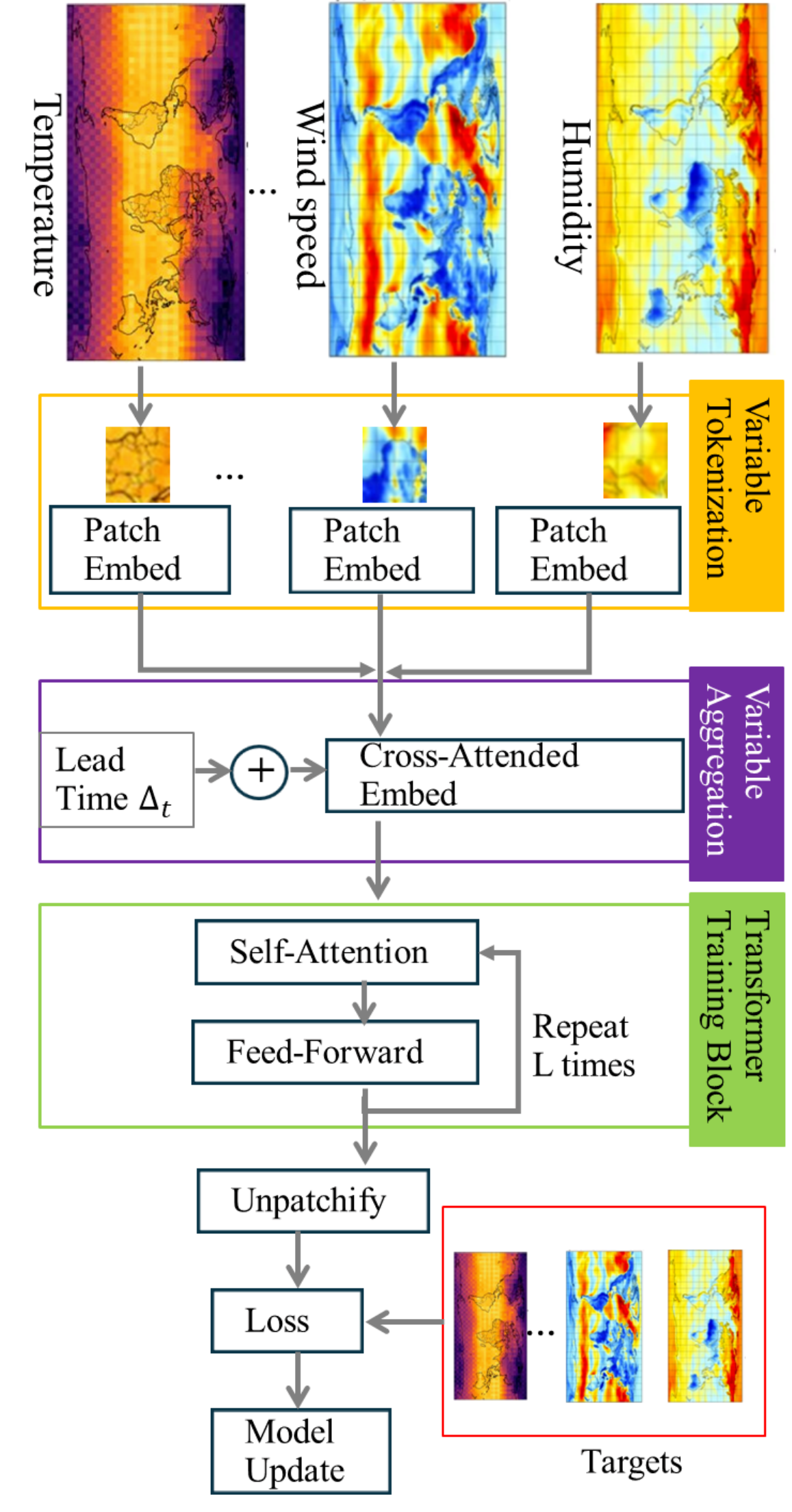}
\caption{Architecture of ClimaX foundational model.}
\label{fig:climax-arch}
\end{figure}

To scale ORBIT, we propose a novel Hybrid Sharded Tensor-Data Orthogonal Parallelism (Hybrid-STOP) by taking advantage of unique mathematical property for matrix chain multiplication and distributing model parameters among GPUs in alternating row and column shards. Through this innovation, the Hybrid-STOP algorithm combines tensor parallelism and fully sharded data parallelism together to achieve better scalability that a single parallelism cannot achieve. Importantly, this approach is designed to be architecture-agnostic, not requiring a specific hardware, which broadens its applicability across various computing platforms.


Our work represents the first instance of pre-training a large ViT model on a non-NVIDIA platform, achieving impressive scaling efficiency and exaFLOPS sustained throughput. This accomplishment is particularly significant given the limited interconnect capabilities of the Frontier platform, especially when compared to the faster interconnects of more recent NVIDIA platforms, and the less advanced software stack typical of non-NVIDIA environments. The successful scaling of the ORBIT model under these conditions not only highlights its robust design but also underscores the potential of alternative hardware platforms for training large AI models, thanks to the innovative Hybrid-STOP scaling approach that we developed.

\section{State of The Art}
\label{sec:SOA}

\begin{figure*}[t]
\begin{subfigure}{.48\linewidth}
\includegraphics[width=\linewidth]{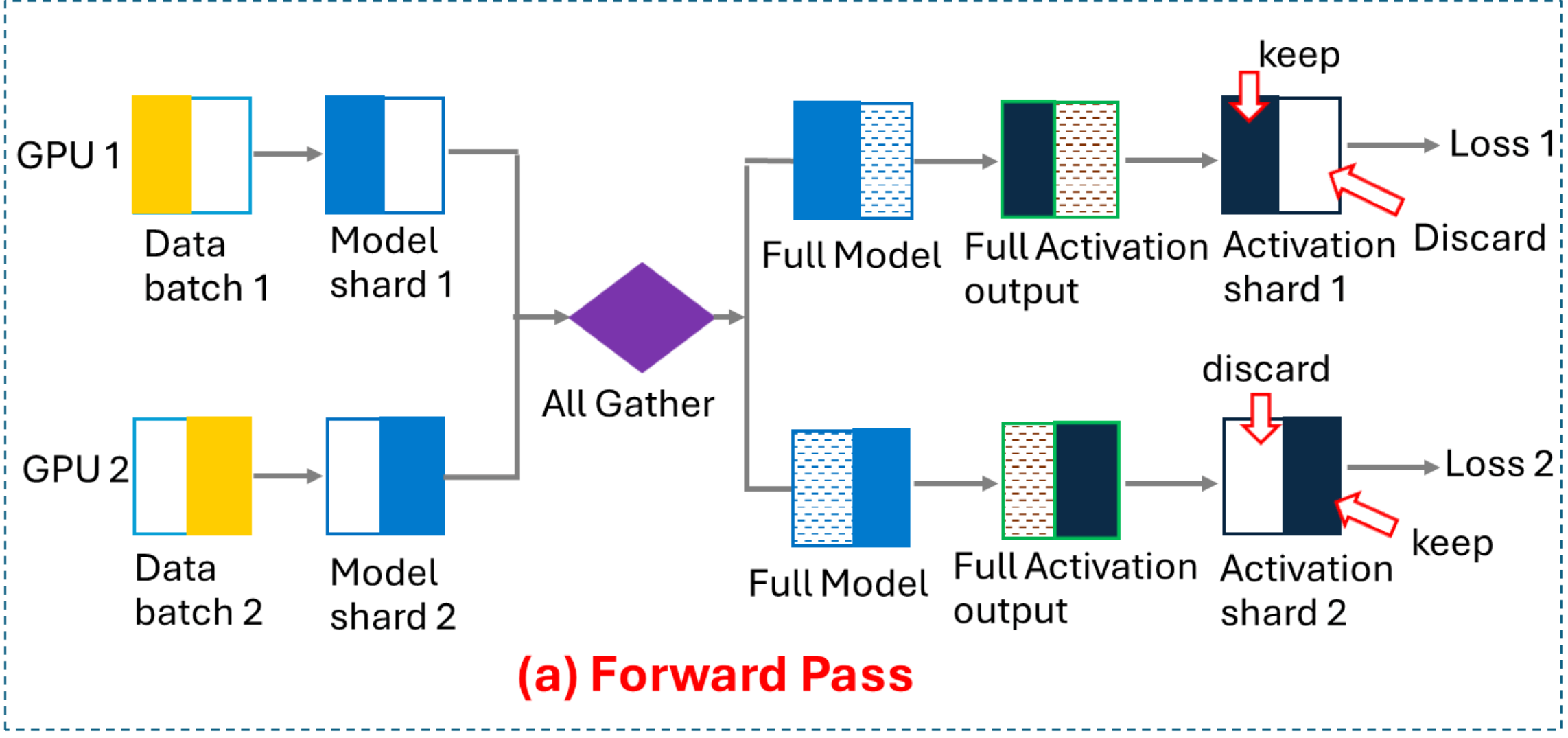}
\end{subfigure}
\begin{subfigure}{.48\linewidth}
\includegraphics[width=\linewidth]{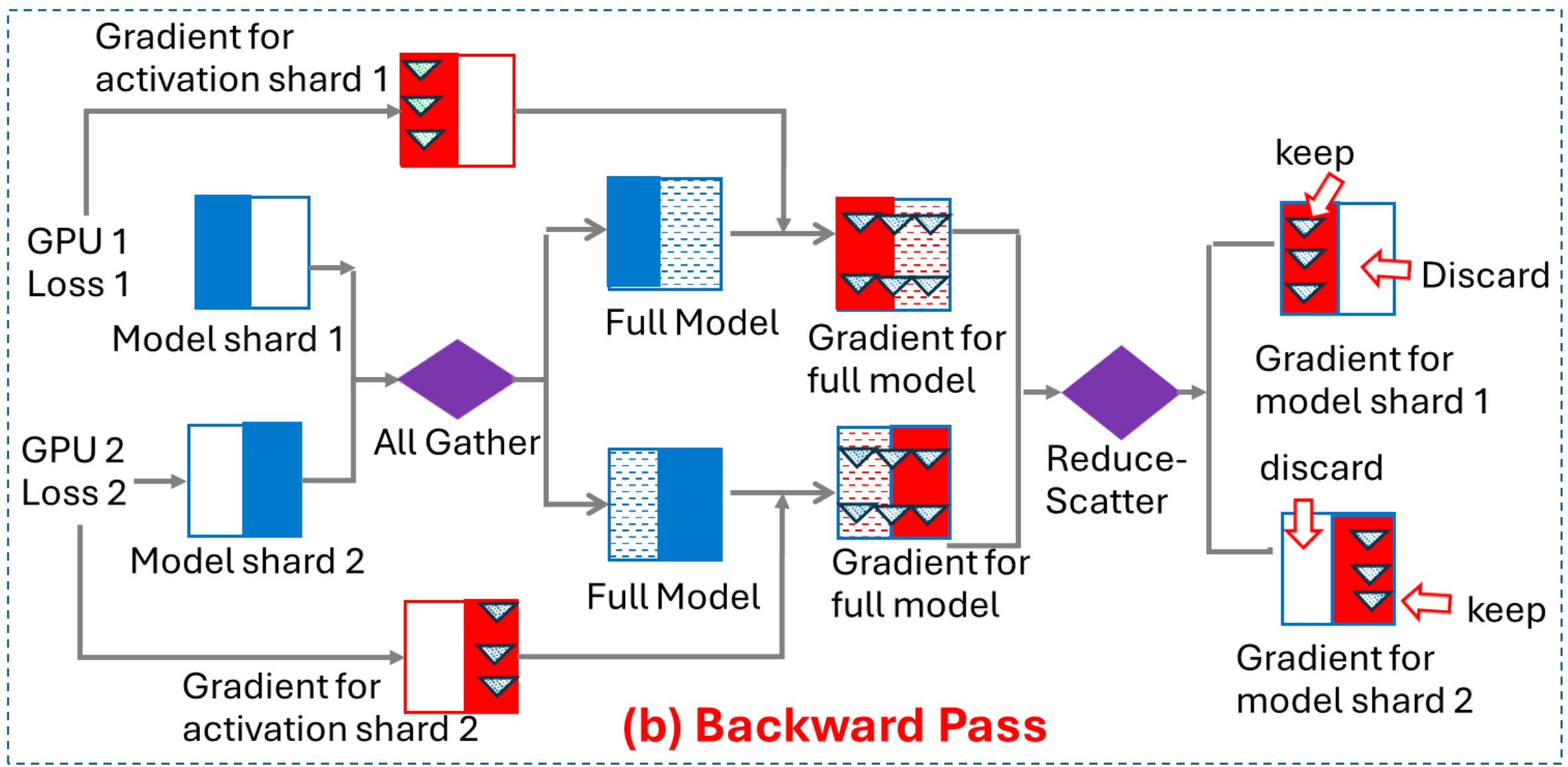}
\end{subfigure}
\caption{Fully Sharded Data Parallelism (FSDP) forward and backward pass.}
\label{fig:fsdp}
\end{figure*}

Fig.~\ref{fig:climax-arch} shows the architecture for the ClimaX foundational model with multiple input channels~\cite{nguyen2023climax}.
Each input channel is a 2D image corresponding to different climate variables, such as humidity, wind speed, and temperature.
After independent patch tokenization for each channel input, the embedding for the tokenized image patches is aggregated together across different channels using cross-attention. The aggregated embeddings are then used in the transformer training block, with each layer of the training block consisting of a self-attention and a feed-forward sub-layers. Then, predictions are computed by projecting the embedding output from the feature space to the image space.

ClimaX only has 115 million model parameters. The small model size, in turn, limits ViT's capability to learn complex data patterns from diverse and chaotic earth system information. In fact, the limitation of small vision models is not unique to climate modeling. The largest foundational dense ViT model to date has 22 billion parameters~\cite{dehghani2023scaling}, and the size trails far behind those of the NLP models, which are now demonstrating capabilities in the trillion parameters scale~\cite{dash2023optimizing, fedus2022switch,bagualu22}.

Scaling large ViT has unique challenges, as mentioned before in Sec.~\ref{sec:problem_overview}. The memory requirement for training a ViT model surpasses that for an NLP model at the same model scale due to the high dimensionality nature of image data. For climate modeling, ViT can have large channel dimensionality (up to 91 channels in our experiments) and each channel represents a unique state of different climate variables. Large input channels are advantageous, leading to more climate variables to be used for prediction. 
Unfortunately, the memory requirements for training ViT models also escalate rapidly with both the image resolution and the number of input channel variables, resulting in significantly more computations and memory use compared to NLP models.

To reduce computations and memory requirements, an effective parallelism technique is the Fully Sharded Data Parallelism (FSDP)~\cite{fsdp, FSDP23}. It enables distributed computations and memory by sharding both data batches and model parameters. Fig.~\ref{fig:fsdp} explains its operations using a 2-GPU example. In Fig.~\ref{fig:fsdp}(a) forward pass, each GPU receives a different data batch and a different model shard. Then, the two GPUs perform an all-gather operation on their model shards to collect the full model in order to compute the full model activations. Since each GPU is only interested in the activations relevant to its assigned model shard, each GPU in this example discards half of the activations, with GPU 1 keeping the left half and GPU 2 keeping the right half. In Fig.~\ref{fig:fsdp}(b) backward pass, the model shards are all-gathered again to compute the gradients for the full model. Note that since each GPU is assigned a different data batch, the magnitude of the full model gradients is also different across GPUs. To synchronize the gradients, the model gradients are averaged together and then scattered to each GPU through a reduce-scatter operation. Finally, half of the model shard gradients are discarded, in the same way as how activation shards are discarded in the forward pass. Note that in Fig.~\ref{fig:fsdp}, a temporary copy of the full model has to be collected during both forward and backward passes. Therefore, the performance of FSDP is limited by its peak memory use when gathering the full model parameters.

Alternative to FSDP, tensor parallelism~\cite{shoeybi2020megatronlm, xu2021gspmd} does not gather model parameters during training. Instead, each GPU keeps model parameters sharded throughout training, but the GPUs are required to communicate frequently on the activations. In contrast, pipeline parallelism~\cite{he2021pipetransformer,huang2019gpipe,dash2023optimizing, kim2020torchgpipe} does not shard model parameters. It partitions a model instance into stages and distributes stages across GPUs, where both activations and gradients are communicated across stage boundaries. Nevertheless, there are two challenges for tensor and pipeline parallelisms. First, the scalability of both tensor and pipeline parallelisms is limited by the specific model architectures. The scalability for tensor parallelism is limited by the number of multi-attention heads~\cite{shoeybi2020megatronlm}, while the scalability for pipeline parallelism is limited by the number of model layers~\cite{ narayanan2021efficient}. Second, vanilla FSDP does not support tensor and pipeline parallelisms to achieve further speedup and memory reduction. Therefore, constructing a solution that can integrate FSDP with these alternative parallelisms can significantly improve the scalability for ViT models.

\section{Innovation Realized}
\label{sec:innovation}

\subsection{Hybrid Sharded Tensor-Data Parallelism}
\label{sec:innovation-hybrid}
To address the challenges discussed in Sec.~\ref{sec:SOA}, we developed a novel Hybrid Sharded Tensor-Data Orthogonal Parallelism (Hybrid-STOP). It takes advantage of unique mathematical property for matrix chain multiplication and divides model parameters in alternate row and column shards. Through this innovation, the Hybrid-STOP algorithm combines the tensor parallelism and FSDP together to achieve better scalability than any single parallelism. In addition, our proposed parallelism technique is a general approach and is not limited by specific model architectures. Furthermore, the Hybrid-STOP algorithm avoids the peak memory use problem as in FSDP and leads to better memory reduction capability, by keeping the parameters sharded throughput training.

To understand the Hybrid-STOP algorithm, let us first explain the key mathematical property that the algorithm utilizes. If we performs two matrix multiplications in order, namely $y\gets xAB$, where $x$, $A$, $B$ are all matrices and $y$ is the output, then the above computations are mathematically equivalent to: 
  \begin{align}
    y \gets xAB &= x[A_{*,1} , A_{*,2}]\begin{bmatrix}
           B_{1,*} \\
           B_{2,*} \\
         \end{bmatrix} \ ,
  \end{align}
where $A_{*,1}$ and $A_{*,2}$ are two sub-matrices of $A$, split along the column direction. $B_{1,*}$ and $B_{2,*}$ are two sub-matrices of $B$, split along the row direction. In a more general term:
  \begin{equation}
y \gets xAB = \sum^{K}_{k=1} xA_{*,k}B_{k,*} \ ,
\label{eqn:forward}
\end{equation}
where $K$ is the number of sub-matrix shards. $A_{*,k}$ is a column shard for matrix $A$ and $B_{k,*}$ is a row shard for matrix $B$. The same property can also be applied to gradient computations. If we denote the gradient of $y$ with respect to $x$ as $\frac{\partial y}{\partial x}$, then:
\begin{equation}
\frac{\partial y}{\partial x} = \left( B^T A^T \right) \otimes I =\left(\sum^{K}_{k=1} B^{T}_{k,*}A^{T}_{*,k} \right) \otimes I \ ,
\label{eqn:backward}
\end{equation}
where $\otimes$ represents the Kronecker product and $I$ is an identity matrix with dimension equal to the number of rows in $x$ and $y$. Notably in this equation, the gradient corresponding to the same row of $y$ and $x$ equals $B^T A^T$, whereas for different rows, the gradient is a zero vector.

The above equations are directly relevant to the core operations of the ViT training block. Each layer of the transformer training block, as shown in Fig.~\ref{fig:climax-arch}, consists of a self-attention and a feed-forward sub-layer. For both sub-layers, the primary computations can be summarized as performing two matrix multiplications in sequence. For the self-attention sub-layer, the primary computations are $\text{softmax}\left( QK^T \right) V$, where $Q$ is a query matrix, $K$ is a key matrix and $V$ is a value matrix~\cite{Vaswani17}. For the feed-forward sub-layer, the primary computations are $\text{GeLU}\left( xA \right)B$, where $x$ is input to the feed-forward sub-layer. $A$ and $B$ are model parameters for linear transformation~\cite{Vaswani17, shoeybi2020megatronlm}. Notice that even if we include the additional computations for softmax and GeLU activation, the key operations for both the self-attention and the feed-forward sub-layers are still conformed to the general form: $y \gets xAB$. The Hybrid-STOP algorithm recognizes this pattern between the feed-forward and self-attention sub-layers. It uses Eqns.~(\ref{eqn:forward}) and~(\ref{eqn:backward}) to distribute model parameters and computations for both self-attention and feed-forward, while also integrating tensor and FSDP parallelisms.

\begin{figure*}[t]
\begin{subfigure}{.4\linewidth}
\includegraphics[width=\linewidth]{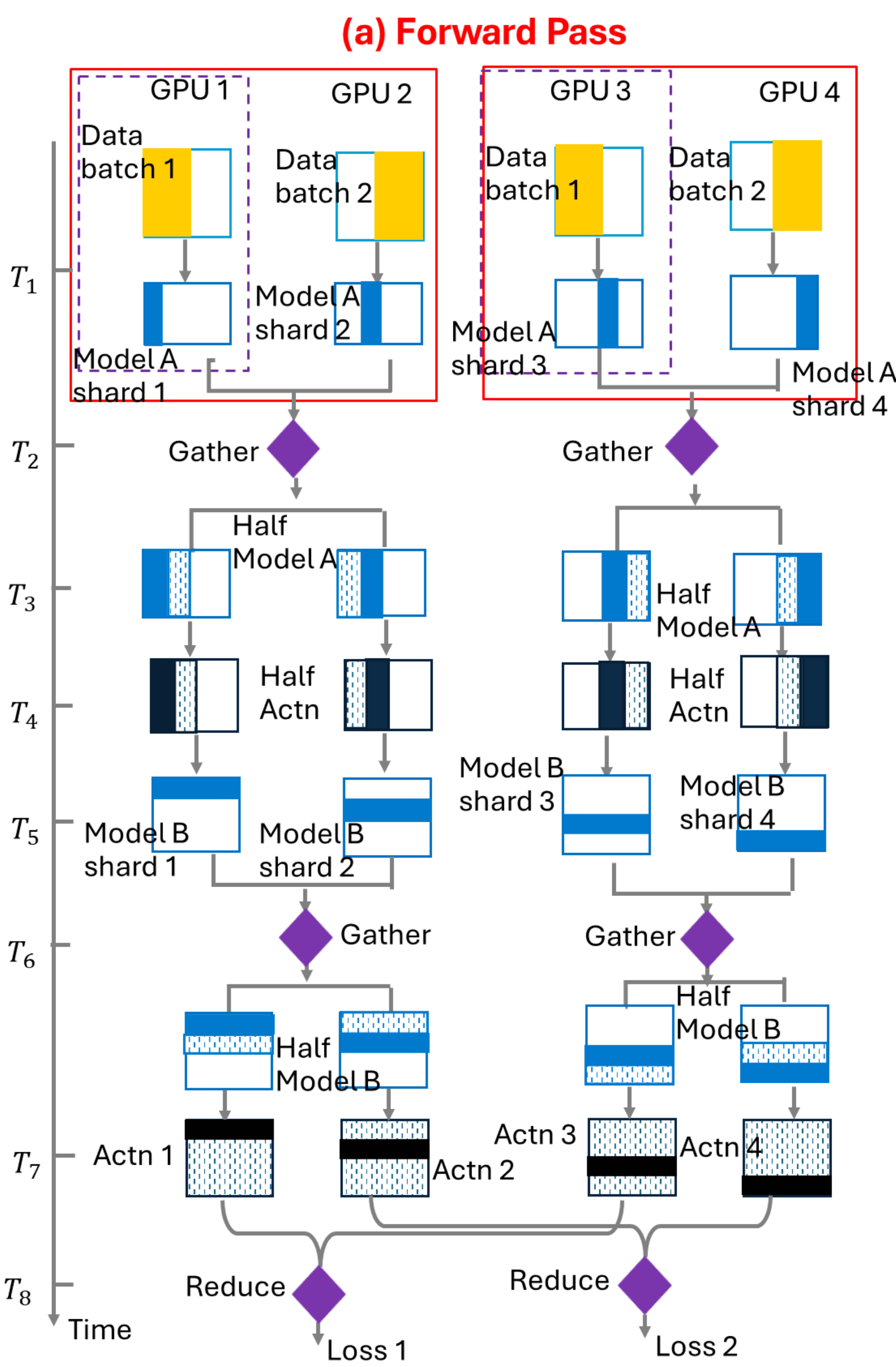}
\end{subfigure}
\hspace{0.1\linewidth}
\begin{subfigure}{.4\linewidth}
\includegraphics[width=\linewidth]{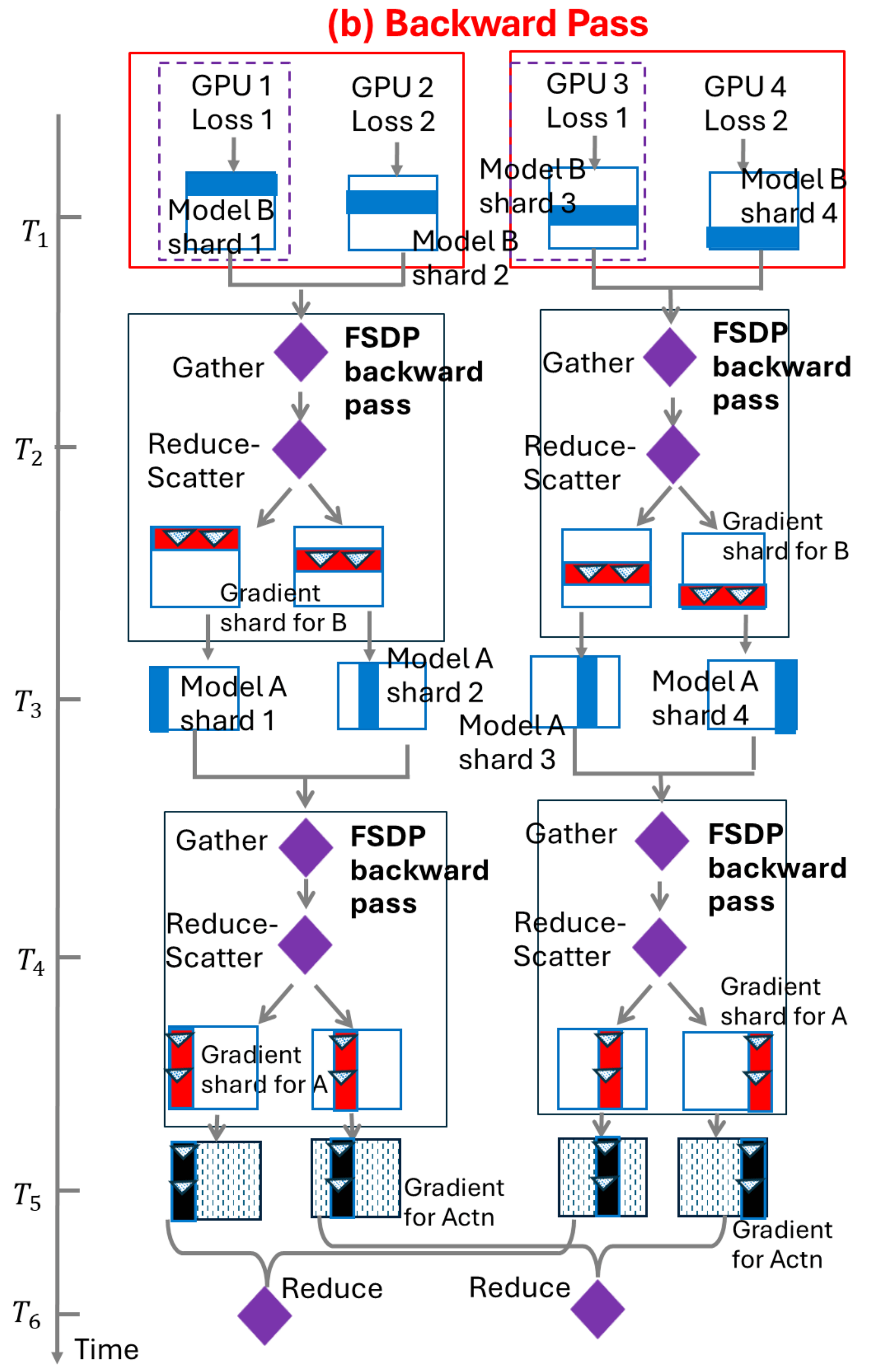}
\end{subfigure}
\caption{Hybrid-STOP forward and backward pass. GPUs 1 and 2 are an example FSDP group, highlighted by a red rectangular box. GPUs 1 and 3 are an example tensor-parallel group, highlighted by purple dash lines.}
\label{fig:hybrid}
\end{figure*}

Fig.~\ref{fig:hybrid} explains the Hybrid-STOP operations using an example with 4 GPUs. Fig.~\ref{fig:hybrid}(a) shows the forward pass. GPUs 1 and 2 are an FSDP group. GPUs 3 and 4 are another FSDP group. Both FSDP groups are outlined by red rectangular boxes and each FSDP group shards both data batches and model parameters. Meanwhile, GPUs 1 and 3 are a tensor-parallel group, highlighted by purple dash lines, and GPUs 2 and 4 are another tensor-parallel group. Each tensor-parallel group shares the same data batches, but shards different model parameters. 

At time steps $T_2$ and $T_3$ in Fig.~\ref{fig:hybrid}(a), the first FSDP group, consisting of GPUs 1 and 2, gathers the left half columns of the model parameters $A$ from Eqns.~(\ref{eqn:forward}) and~(\ref{eqn:backward}). Then the activations corresponding to the left half columns are computed at time step $T_4$. Similarly, the second FSDP group, consisting of GPUs 3 and 4, gathers the right half column shard of the model parameters $A$ and computes half activations. Then at time step $T_6$, the first FSDP group gathers the model shards for the top half row shard for the model parameters $B$ from Eqns.~(\ref{eqn:forward}) and~(\ref{eqn:backward}), and the second FSDP group gathers the bottom half row shard for $B$. By multiplying the half columns activation from time step $T_4$ with the half row shard for $B$ from time step $T_6$, each GPU computes an individual activation output, $xA_{*,k}B_{k,*}$ at time step $T_7$. Then GPUs in each tensor-parallel group add their individual activations together, based on the mathematical principle defined in Eqn.~(\ref{eqn:forward}), and compute the final activation and training loss.

In the backward pass in Fig.~\ref{fig:hybrid}(b), each FSDP group gathers  their row shards for model parameters $B$ at time step $T_1$. Then each FSDP group performs a gather and a reduce-scatter operations to compute the gradients for the parameter row shards at time step $T_2$, in the same manner as the FSDP algorithm operates in Fig.~\ref{fig:fsdp}(b). The only difference is that hybrid-STOP does not gather the full model. Instead, GPUs 1 and 2 gathers the first half row shard, and GPUs 3 and 4 gathers the second half row shard. Then, each FSDP group gathers column shards of model parameters $A$ at step $T_3$ and compute the gradient for its column shard at step $T_4$. At time step $T_5$, the gradients for each GPU's activation are computed. To synchronize gradients, GPUs in the same tensor-parallel group add their activation gradients together through all-reduce and then perform the Kronecker product with an identity matrix to compute the final gradients, following the mathematical principle in Eqn.~(\ref{eqn:backward}).

Note that in Fig.~\ref{fig:hybrid}, the Hybrid-STOP does not gather a temporary copy of the full model like the FSDP algorithm does in Fig.~\ref{fig:fsdp}.
Instead, the Hybrid-STOP keeps the model sharded, thereby leading to lower peak memory footprint and scaling to larger model size.
In addition, since the proposed Hybrid-STOP algorithm is based on the general form of matrix chain multiplications in Eqns.~(\ref{eqn:forward}) and~(\ref{eqn:backward}), Hybrid-STOP can be applied to for both feed-forward and self-attention, irrespective to transformer architecture. This is in contrast to tensor parallelism, whose scalability is limited by the number of self-attention heads, or pipeline parallelism, whose scalability is limited by the number of layers.

\subsection{Optimization}
\label{sec:optimization}
This section discusses various optimization techniques that were used to aid training convergence, scalability, computing efficiency and memory use.

\noindent\ul{\textbf{Architecture Optimization.}} Prior work on training 22-billion ViT model reports divergent training loss due to extremely large attention logits with near-zero entropy~\cite{dehghani2023scaling}. To solve this issue, we adopted the same approach from~\cite{dehghani2023scaling}, by applying additional layer normalization to the queries, $Q$, and keys, $V$, before computing the self-attention scaled dot product. Through these additional layer normalization, the attention logit growth is contained and the training loss divergence is prevented.

\begin{figure}[t]
\centering
\includegraphics[width=.9\linewidth]{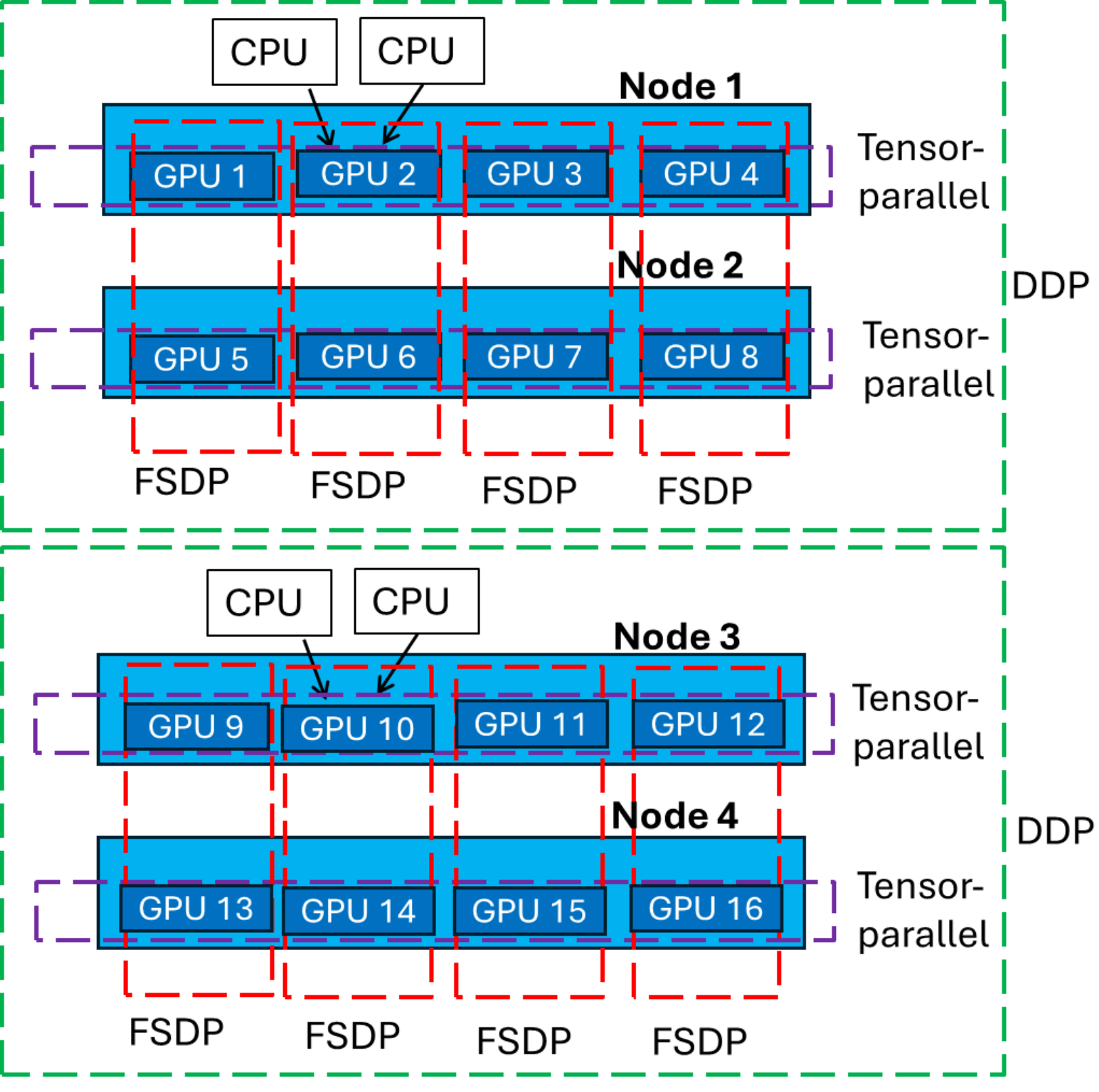}
\caption{Hierarchical parallelism of the Hybrid-STOP. Each horizontal purple rectangle represents a tensor-parallel group. Vertical red rectangles represent FSDP groups. Green rectangles represent DDP groups.}
\label{fig:hierarchical-parallel}
\end{figure}

\noindent\ul{\textbf{Hierarchical Parallelism.}} 
Integrating the FSDP and tensor parallelisms does not provide a sufficient source of parallelism to scale to a large supercomputer. To address that, additional parallelism is needed.  In addition, how to map the Hybrid-STOP algorithm to the architecture of a large supercomputer also requires special consideration.

To scale to large supercomputer, we add Distributed Data Parallelism (DDP)~\cite{li2020pytorch} to Hybrid-STOP as an orthogonal level of parallelism. Fig.~\ref{fig:hierarchical-parallel} explains the interactions among different parallel groups and how they are mapped to the hierarchical structure of a supercomputer. The tensor parallelism requires frequent fine-grain communications to perform activation reduction in each training layer. Therefore, GPUs in the same tensor-parallel group are mapped to GPUs in the same node of a supercomputer to take advantage of low latency peer-to-peer network in a node. For illustration, the tensor-parallel groups are highlighted by horizontal purple rectangles in Fig.~\ref{fig:hierarchical-parallel}. Meanwhile, each GPU is connected with multiple CPUs to load input data across different climate variable channels.

In contrast, FSDP parallel groups are mapped to GPUs from different nodes, which has slower node-to-node network than the tensor-parallel group's peer-to-peer network within the same node. We design it this way because the FSDP parallel group requires coarser communication by performing gather and reduce-scatter communications on a model shard. In the example of Fig.~\ref{fig:hierarchical-parallel}, GPUs 1 and 5 is an FSDP group and is highlighted by a verticle red rectangle. GPUs 2 and 6, 3 and 7, 4 and 8 are other FSDP groups.

The DDP groups require least communication and they require only one gradient reduction for each global data batch. Therefore, each DDP group is mapped to a sub-cluster of the supercomputer and each sub-cluster consists of multiple nodes. In addition, different DDP groups are assigned with different data subsets to train. In the example of Fig.~\ref{fig:hierarchical-parallel}, nodes 1 and 2 form a DDP group, and nodes 3 and 4 is another DDP group. Each DDP group consists of two tensor-parallel groups and four orthogonal FSDP groups.

\noindent\ul{\textbf{Activation Checkpointing.}} We used PyTorch activation checkpointing technique~\cite{gruslys2016memoryefficient, chen2016training} to trade compute for memory saving on large models. Instead of keeping activation outputs from the forward pass in memory for the expected use in backward pass, the activation checkpointing technique omits saving them in memory. Then during the backward pass, the needed activation outputs are recomputed by rerunning the forward pass segment for the activation outputs.

\noindent\ul{\textbf{Mixed-Precision.}} We used BFLOAT16 mixed-precision for faster computation. One issue with using BFLOAT16 mixed-precision is that gradients with small or large magnitudes are not always representable in BFLOAT16, and are either flushed down to zero or exploded to infinity. To address the above mentioned issues, we applied the PyTorch dynamic gradient scaling mechanism~\cite{pytorch-gradscaling} during our training. It automatically detects the value range of the computed gradient magnitude. If the values are beyond the value range of BFLOAT16, the gradient scaling will scale the gradient magnitudes to the proper value range representable in BFLOAT16, and then reverse the scaling during the parameters update.

\noindent\ul{\textbf{Layer Wrapping and Prefetching.}} To further reduce communication overhead, the Hybrid-STOP does not shard model parameters from all training block layers at once. Instead, it shards model parameters layer by layer to reduce the size of communication messages. At each layer, the FSDP groups communicate on shards of the model parameters from the same single layer each time before moving on to the next layer. 

In addition, the Hybrid-STOP algorithm prefetches the gathering for the model shards before the model shards are needed, so that cross-GPU communication can occur asynchronously and the communication overhead can be hidden in the computations. For example in Fig.~\ref{fig:hybrid}(a) forward pass,  the gathering for the row shards of model parameter $B$ is needed at time step $T_6$. However, the Hybrid-STOP algorithm can prefetch the gathering when computing the activations at time step $T_4$, thereby hiding the gathering communication overhead in the computations. Similarly in the backward pass in Fig.~\ref{fig:hybrid}(b), the gathering for the model column shards at time step $T_4$ can be prefetched at time step $T_2$.

\begin{figure}
\centering
\includegraphics[width=.9\linewidth]{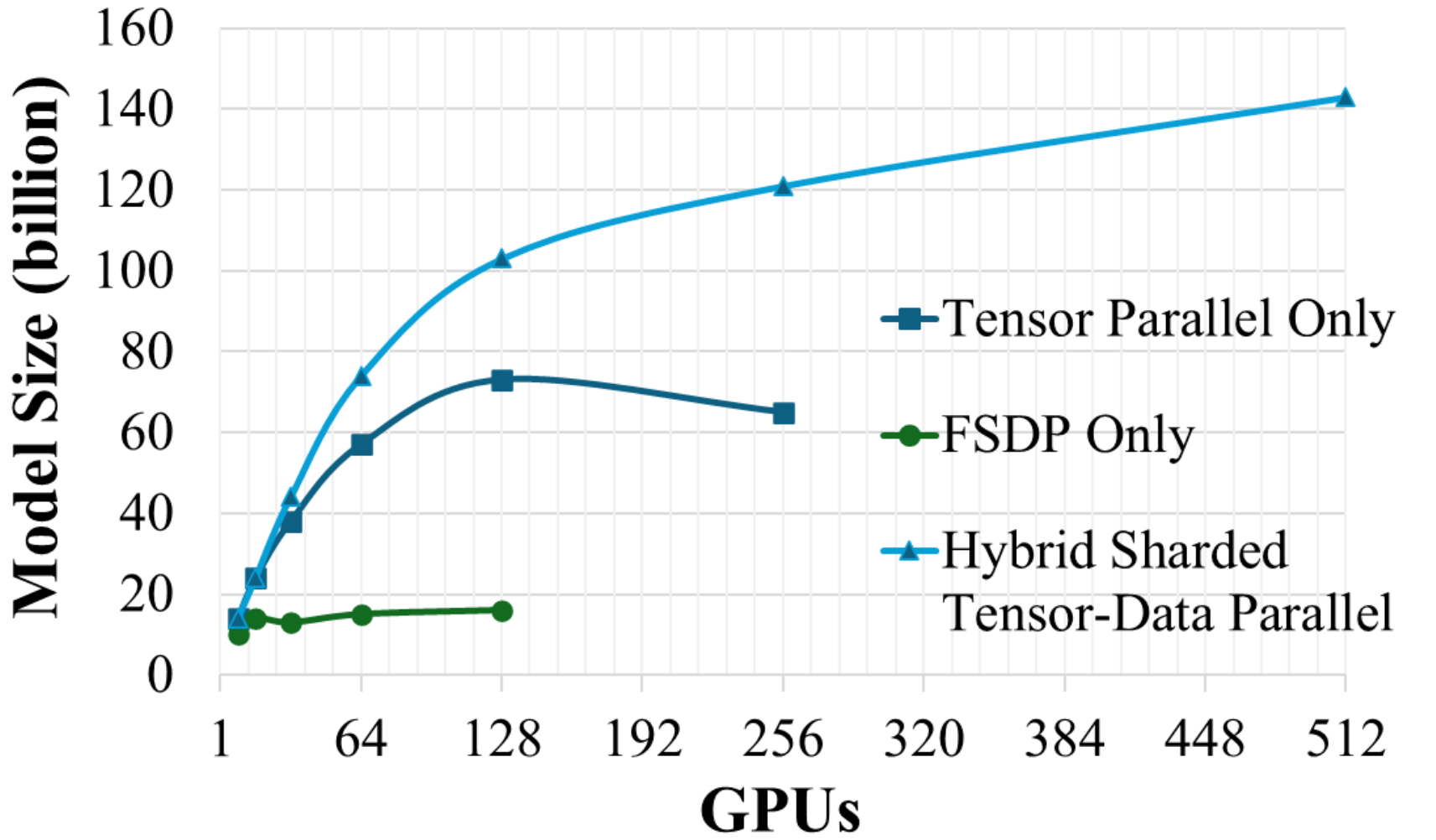}
    \caption{The maximal model size that each parallelism can scale to at different numbers of GPUs.}
\label{fig:maximal_model_size}
\end{figure}

\section{How Performance Was Measured}
\label{sec:experiment setup}
\noindent\ul{\textbf{Model Configuration.}}
All scalability performance numbers presented in the Result Sec.~\ref{sec:performance_result} are based on five different ViT model sizes: 115 million (1024 embedding, 8 layers, 16 attention heads), 1 billion (3072 embedding, 8 layers, 16 attention heads), 10 billion (8192 embedding size, 11 layers, 32 attention heads) and 113 billion parameters (12288 embedding size, 56 layers, 64 attention heads). The architecture was the same as the ClimaX AI foundational model~\cite{nguyen2023climax}, with a figure illustration in Fig.~\ref{fig:climax-arch}. The only exception is that we used additional layer normalization to self-attention queries and key vectors, as described before in Sec.~\ref{sec:optimization}, to prevent training loss divergence.

\noindent\ul{\textbf{Pre-training Dataset.}}
We used pre-training data from CMIP6~\cite{cmip}, which is an international effort across different  climate modeling groups to compare and evaluate climate models. The data can be accessed from the CMIP6 archive~\cite{cmip6-archive}. We used ten different pre-training data sources from CMIP6 (MPI-ESM, AWI-ESM, HAMMOZ, CMCC, TAI-ESM, NOR, EC, MIRO, MRI, and NESM). These sources provide a dataset spanning 65 to 100 years of simulation, yielding more than 1.2 million observation data points with a time difference of 6 hours between two consecutive observation data points.

\begin{table}
\centering
\caption{113 billion model walltime per observation data point using 512 GPUs, with and without optimizations from Sec.~\ref{sec:optimization}.}
{\small
\begin{tabular}{l|c|c|c|c|r}
\toprule

Layer Wrapping & $\times$ & \checkmark & \checkmark  & \checkmark & \checkmark\\
Mixed Precision & $\times$ & $\times$ & \checkmark  & \checkmark  & \checkmark\\
Prefetching & $\times$ & $\times$ & $\times$ & \checkmark & \checkmark\\
Activation Checkpoint & $\times$ & $\times$ & $\times$ & $\times$ & \checkmark\\
\midrule
Walltime (secs) & OOM & 0.97 & 0.49 & 0.40 &0.17\\
\bottomrule
\end{tabular}
}
\label{table:optimization}
\end{table}

Each pre-training observation data point has size $128 \times 256 \times C_d$, where $128 \times 256$ is an image for a single climate variable with a spatial resolution of 1.40625 degree. $C_d$ is the configurable input channel dimension and represents the number of climate variables to be used for training.
At each model size (115 million, 1 billion, 10 billion, and 113 billion), we pre-trained two different ViT models, one with an input channel of 48 climate variables ($C_d=48$) and one with an input channel of 91 variables ($C_d=91$). The 48 climate variables are chosen based on the variable settings in ClimaX~\cite{nguyen2023climax}. The 91 variables are carefully chosen by climate scientists to represent an even broader range of physical variables, including 3 static variables, 3 surface variables, and 85 atmospheric variables spanning 17 pressure levels in the atmosphere. The full list of the 91 chosen variables will be provided to readers when the models are released on GitHub.

\begin{figure*}
\centering
\begin{subfigure}{.48\linewidth}
\includegraphics[width=\linewidth]{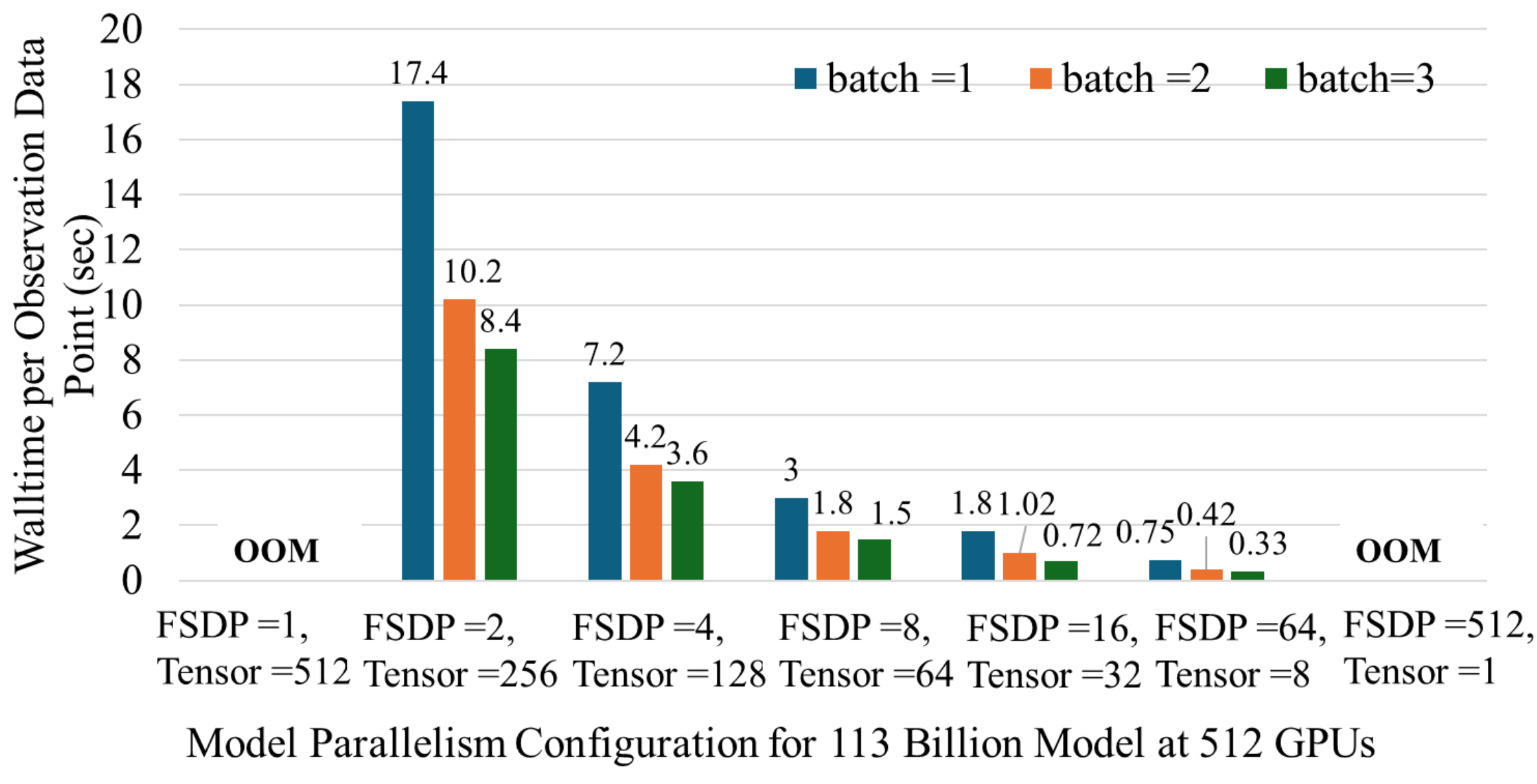}
    \caption{}
\end{subfigure}
\begin{subfigure}{.48\linewidth}
\includegraphics[width=\linewidth]{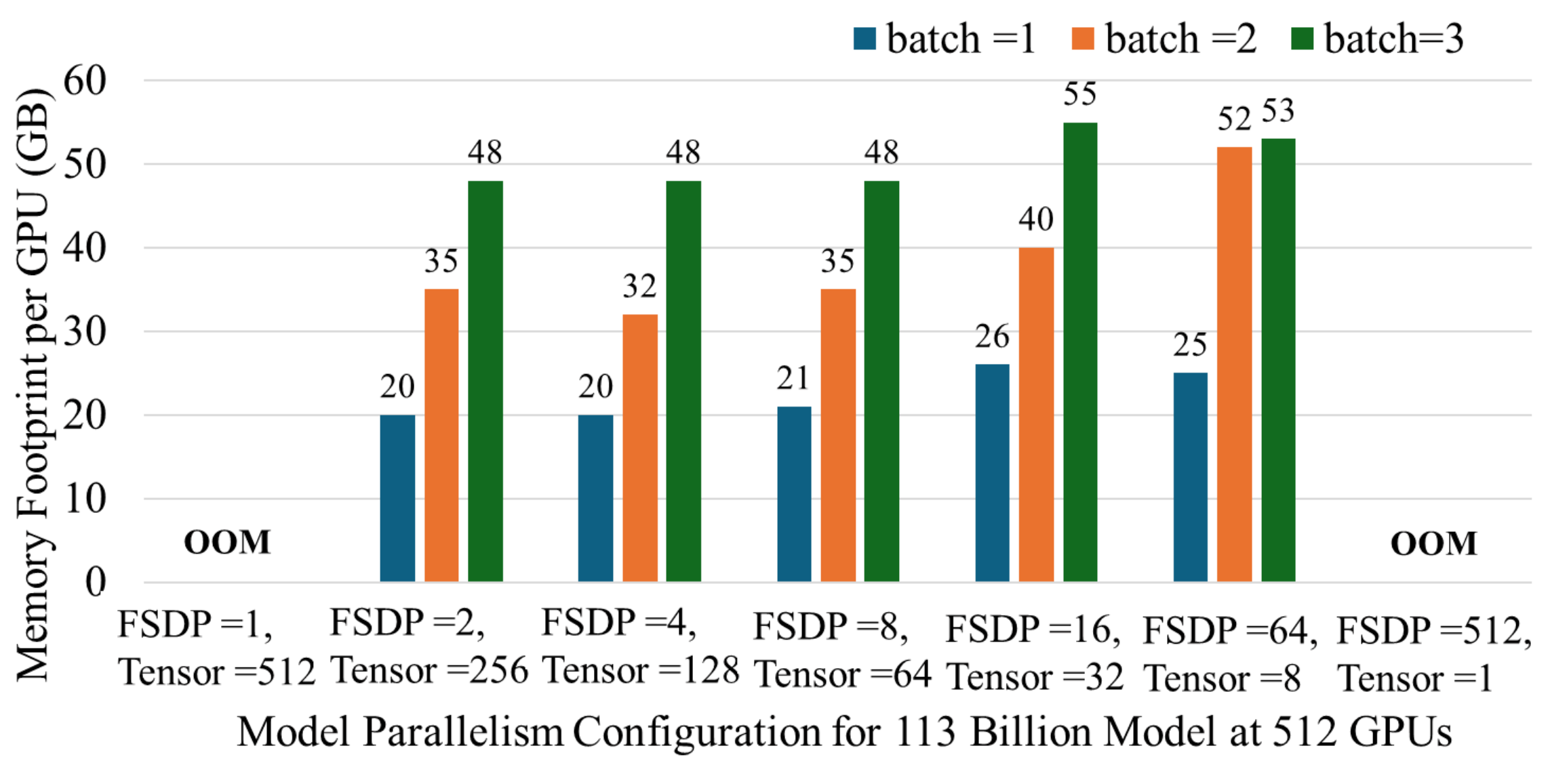}
    \caption{}
\end{subfigure}
\caption{Time-to-solution in (a) and memory footprint in (b) at 512 GPUs with different parallelism configurations.}
\label{fig:parallel-configuration}
\end{figure*}

\noindent\ul{\textbf{Fine-Tuning Dataset.}}
We utilized the 1.40625-degree ERA5 dataset~\cite{era5}, spanning from 1979 to 2020, for model fine-tuning. Specifically, data from 1979 to 2018 was allocated for training the fine-tuning data. Data from 2019 and 2020 was used for validation and evaluation, respectively. This partition of the fine-tuning dataset aligns with the current standard used by the Weatherbench2~\cite{rasp2023weatherbench}. The fine-tuning dataset has the same spatial resolution as pre-training and uses 91 variables, with an input observation data point size of $128 \times 256 \times 91$. The output variables selected were geopotential at 500 hPa (z500), temperature at 850 hPa (t850), 2-meter temperature (t2m), and zonal wind at 10 meters (u10).

\noindent\ul{\textbf{System Details.}}
Experiments were performed using the Frontier Supercomputer at the Oak Ridge Leadership Computing Facility. Each Frontier node has a single 64-core AMD EPYC CPU and 8 GPUs, with 64 GB of memory for each GPU. Among the 8 GPUs, every two of them share an MI250X graphics card and are connected with Infinity Fabric CPU-GPU.
The four MI250X graphics card in the same node are connected with Infinity Fabric GPU-GPU of 50GB/s. The nodes are connected via a Slingshot-11 interconnect with 100GB/s. For the software, we used PyTorch v2.2, ROCm v5.7.0, MIOpen v2.19.0, RCCL v2.17 with libfabric v1.15.2 plugin.

\noindent\ul{\textbf{Performance Metrics.}}
The total number of floating point operations (FLOPs) of the systems was collected via the Microsoft Deepspeed Profiler~\cite{deepspeed} and we only gathered the FLOPs on GPUs. Only the mixed-precision BFLOAT16 results were reported. The scalability of the Hybrid-STOP algorithm was measured on the pre-training datasets using the following three metrics:
\begin{itemize}[leftmargin=*]
\item {\textit{Time-to-solutions}}. We reported two time-to-solutions for pre-training, one for 48 channel variables and one for 91 channel variables. The size of each 48-channel observation data point is $128 \times 256 \times 48$. The size of each 91-channel data point is $128 \times 256 \times 91$. The time-to-solution is defined as the average wall clock time to process each observation data point. The time-to-solution also equals to the runtime per epoch, divided by the number of data points in an epoch.
\item {\textit{Strong scaling efficiency}}. Defined as the speedup for training an epoch at different numbers of GPUs, divided by the number of GPUs. We use the runtime at 512 GPUs (64 nodes) as the 100\% efficiency baseline.
\item {\textit{Training loss}}. We provided the pre-training loss for all 4 model sizes, with the training loss computed as the latitude-weighted Mean Squared Error (wMSE).
\end{itemize}


For the final model prediction performances on fine-tuning evaluation data, predictions were evaluated over the entire grid points for the year 2020. The area differences in the grid cells from the poles to the equator yield an inordinate bias to the polar regions, if all cells are weighed equally. Thus, we evaluated the prediction performance based on latitude-weighted Anomaly Correlation Coefficient (wACC). wACC is calculated as the Pearson correlation coefficient of the anomalies with respect to the climatology for each variable, with the values ranging from -1 (perfect anti-correlation) to 1 (perfect correlation). Value of 0 represents predictions identical to the climatology. A higher wACC suggests that the model predictions closely follow the actual spatial variations observed in climate data, which is desirable.

\begin{figure*}
\centering
\begin{subfigure}{.49\linewidth}
\includegraphics[width=\linewidth]{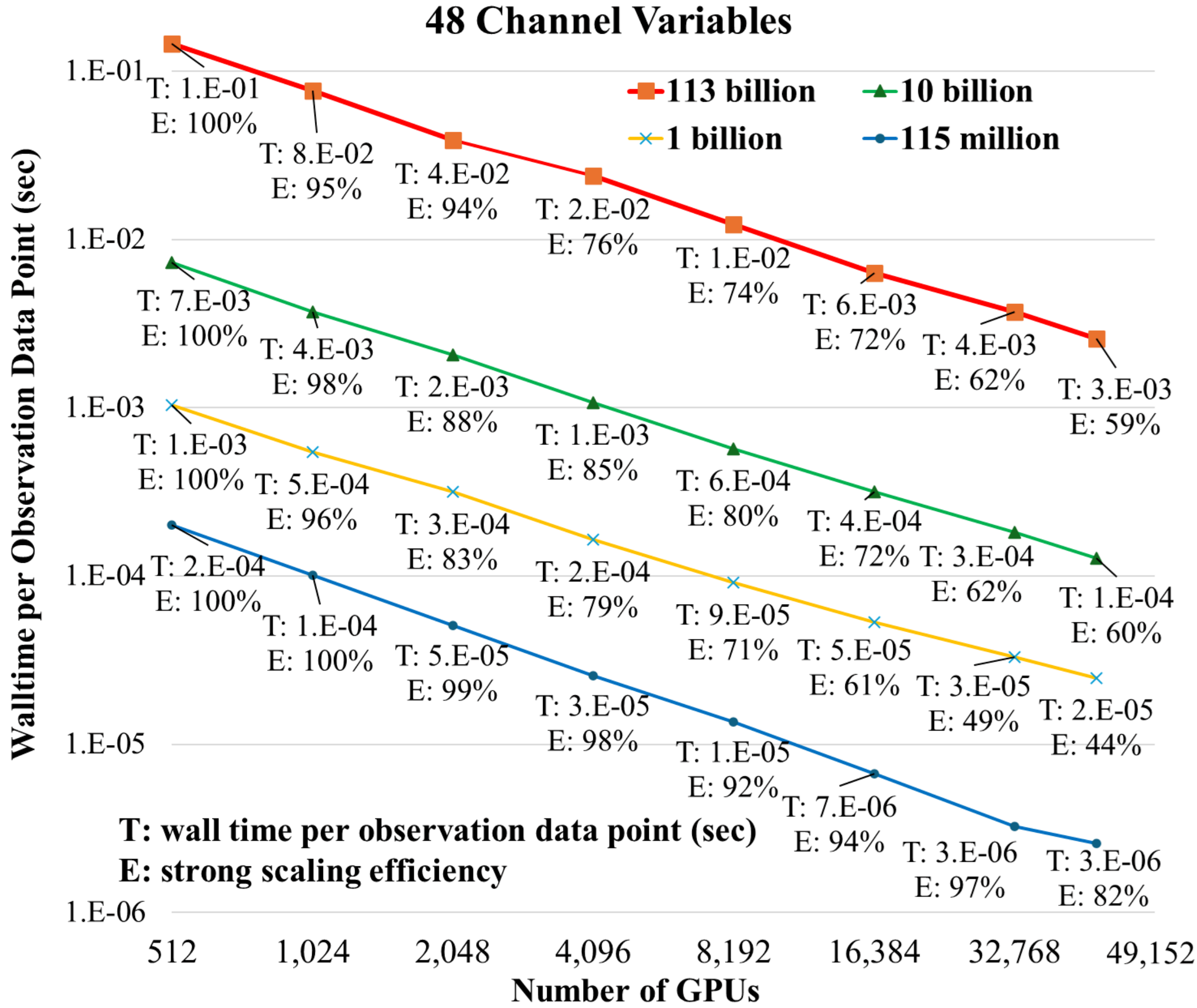}
    \caption{Strong Scaling for 48 Climate Variables}
\end{subfigure}
\begin{subfigure}{.49\linewidth}
\includegraphics[width=\linewidth]{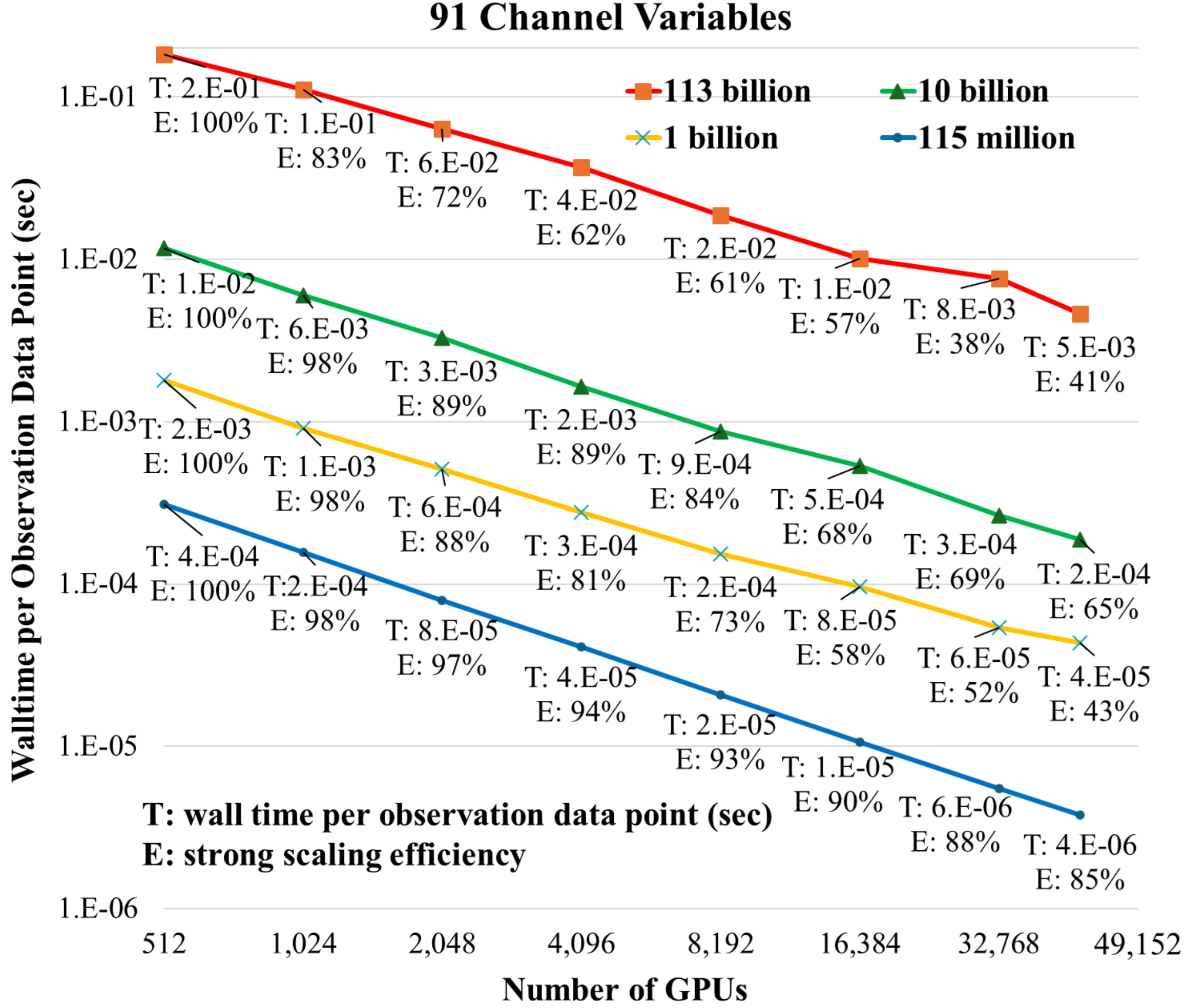}
    \caption{Strong Scaling for 91 Climate Variables}
\end{subfigure}
\caption{The strong scaling efficiencies and time-to-solutions for different model sizes, with 48 channels of climate variables in (a), and with 91 channels of variables in (b). $T$ represents the walltime to process each observation data point. $E$ represents the strong scaling efficiencies, compared to runtimes at 512 GPUs.}
\label{fig:strong-scaling}
\end{figure*}

\section{Performance Results}
\label{sec:performance_result}
\subsection{Scaling Maximal Model Size}
Fig.~\ref{fig:maximal_model_size} shows the maximal model sizes that tensor parallelism, FSDP, and the proposed Hybrid-STOP can scale to from 1 to 512 GPUs (64 nodes). All runs in the figure used batch size of 2 and 48 channels of climate variables. Note that the model size for FSDP can only be scaled to 20 billion parameters. The model size for tensor parallelism is scaled to 73 billion parameters. The Hybrid-STOP algorithm, however, scales the model size to 143 billion parameters at 512 GPUs. Both FSDP and tensor parallelism have their limitations. The model size scaling performance for FSDP is limited by its peak memory use when collecting temporary copies of model parameters. Tensor parallelism, however, is limited by the ViT architecture as the amount of tensor parallelism cannot be more than the number of attention heads. In comparison, the hybrid-STOP does not have either limitation and enable better model scaling capability. 

\subsection{Optimization Techniques}
Table~\ref{table:optimization} shows the walltime comparison, with and without the optimizations techniques from Sec.~\ref{sec:optimization}. All reported numbers in the table were for the 113 billion parameters model on 512 GPUs (64 nodes), using 48 channels of variables. The program was out of memory (OOM) when none of the optimization techniques was used. With layer wrapping alone, walltime was 0.97 seconds per observation data point. With both layer wrapping and BFLOAT16 mixed-precision, walltime was reduced to 0.49 seconds. With all four optimizations, the walltime was further reduced to 0.14 seconds.

\subsection{Configurations for Hierarchical Parallelism}
The Hybrid-STOP algorithm uses three orthogonal parallelism groups, which are FSDP, tensor, and DDP. The configuration of the parallel group size for each level of parallelism has a noticeable impact on both runtimes and memory use. Fig.~\ref{fig:parallel-configuration} shows the walltime and memory use for the 113 billion model to process each observation data point. All numbers in the figure were run on 512 GPUs and used a DDP group size of 1. Performance numbers were presented with different combinations of FSDP and tensor parallel group sizes. The program ran out of memory when using either FSDP or tensor parallelism alone. Using an FSDP group size of 64 and a tensor parallel group size of 8, the program has the fastest computation speed  with 0.33 walltime second per data point when using a batch size of 3.  This runtime is 25 times faster than the performance when using a FSDP group size of 2 and a tensor parallel group size of 256. The hierarchical parallelism configurations, however, have less impact on memory use. A mild memory footprint increase can be observed when increasing the FSDP and decreasing the tensor parallel group size.

\subsection{Strong Scaling and Time-to-Solutions}
Fig.~\ref{fig:strong-scaling} shows the strong scaling efficiencies, denoted as ``E" in the figure, and walltime per observation data points, denoted as ``T" in the figure, from 512 GPUs (64 nodes) to 49,152 GPUs (6,144 nodes). Experiments in Fig.~\ref{fig:strong-scaling}(a) used 48 channels of climate variables and experiments in Fig.~\ref{fig:strong-scaling}(b) used 91 channels of climate variables. We observed that the Hybrid-STOP algorithm retained excellent strong scaling efficiency when scaling up the model size and the input channel size. With 48 input channels, all four model sizes (113 billion, 10 billion, 1 billion, and 115 million parameters) maintained strong scaling efficiencies in the range of 44\% to 82\% at 49,152 GPUs, compared to the baseline performance at 512 GPUs. For the performance with 91 input channels, all model sizes retained strong scaling efficiencies in the range of 41\% to 85\% at 49,152 GPUs.

For time-to-solutions, the 113 billion parameter model takes 3E-03 second for each 48-variable observation data point at 49,152 GPUs with a sustained computing throughput at 684 PFLOPS. One pre-training epoch takes 0.8 wall-clock hours on 49,152 GPUs for the 113 billion parameter model to see all 1.2 million observation data points. The 10 billion parameter model takes 1E-04 second for each 48-channel observation data point at 49,152 GPUs, with a sustained computing throughput at 1.6 exaFLOPS. In contrast, due to increased memory use for more input channels, processing each 91-channel observation data point takes more walltime, requiring 5E-03 second for the 113 billion parameter model, and 2E-04 second for the 10 billion parameter model at 49,152 GPUs. 

\begin{figure}
\centering
\includegraphics[width=.9\linewidth]{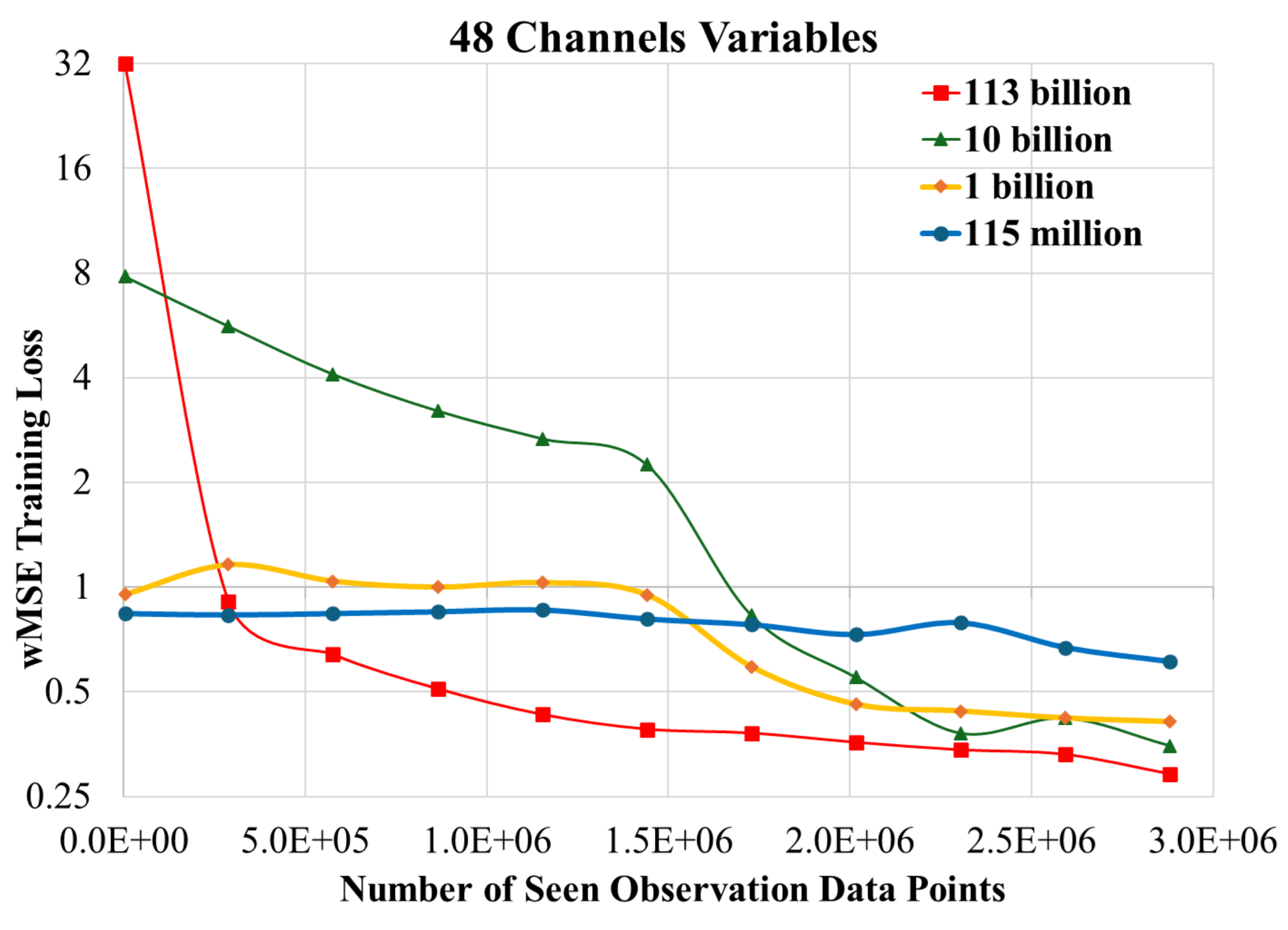}
    \caption{The pre-training loss comparison among different model sizes with 48 input channels.}
\label{fig:training_loss}
\end{figure}

\subsection{Pre-Training Loss}
Pre-training loss for all four model sizes is presented in Fig.~\ref{fig:training_loss}. All four models used a fixed global data batch size of 2,880 and trained for 2.5 epochs. Despite of high initial loss, the models with 10 billion and 113 billion parameters demonstrated greater training data efficiency and converged more rapidly, outperforming the smaller models with 115 million and 1 billion parameters after processing 2 million observation data points.

\subsection{Fine-Tuning Model Prediction}


\begin{figure*}
\centering
    \includegraphics[width=0.7\textwidth]{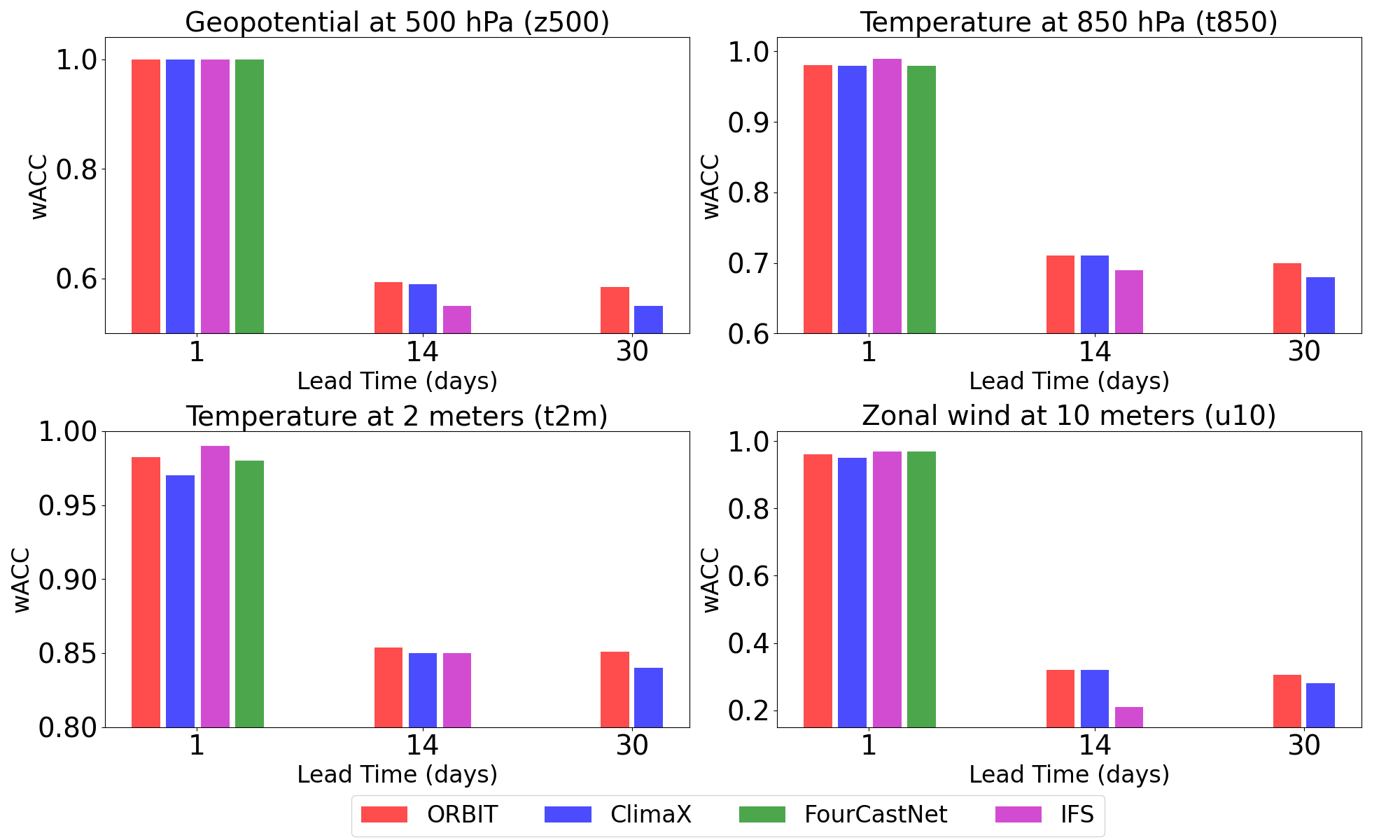}
    \caption{Latitude-Weighted Anomaly Correlation Coefficient (wACC) scores for ORBIT (115 million parameters), ClimaX (same model size), 
    FourCastNet, and IFS models in forecasting four atmospheric variables at 1, 14, and 30-day intervals for the testing year 2020.}
    \label{fig:wACC_115M}
\end{figure*}

Fig.~\ref{fig:wACC_115M} illustrates the performance of the 115 million-parameter ORBIT model in forecasting four key atmospheric variables at 1, 14, and 30-day intervals using the ERA5 fine-tuning dataset. This performance is compared with ClimaX~\cite{nguyen2023climax}, 
FourCastNet~\cite{fourcastnet}, and IFS~\cite{ifs} using the wACC accuracy score. ClimaX is an AI foundation model providing forecasts up to 30 days, and its wACC performance scores were obtained from Fig.~6 of the ClimaX paper~\cite{nguyen2023climax}. In contrast, FourCastNet is not foundation model, but is task-specific AI models trained on ERA5 data with 0.25 degree spatial resolution.
The wACC performance scores were obtained from 
Fig.~6 of the FourCastNet paper~\cite{fourcastnet}. We did not compare with Stormer~\cite{stormer} since Stormer used different climatology data and a direct comparison is not available. IFS is a numerical model from the European Center for Medium-Range Weather Forecasting, with its wACC performance scores obtained from Fig.~6 of the ClimaX paper~\cite{nguyen2023climax}. Notably, both ClimaX and ORBIT provides forecasts up to 30 days, while FourCastNet does not offer either 14 or 30-day forecasts.

Fig.~\ref{fig:wACC_115M} highlights the strength of ORBIT, especially evident at longer lead times, where it consistently excels and outperforms ClimaX
and IFS. For a 14-day forecast, ORBIT shows improvements of up to 52\% over IFS. 
For the 30-day forecast, ORBIT performs up to 9\% better than ClimaX. This superior performance at extended lead times highlights ORBIT's robustness and reliability in long-term forecasting. For 1-day forecasts, ORBIT's accuracy is comparable to that of IFS, FourCastNet, and ClimaX on almost all variables except u10. ORBIT shows a 0.1\% better wACC on t850 prediction compared to ClimaX and FourCastNet. It also shows a 0.27\%-1.30\% improvement in t2m predictions compared to all others, except IFS. For u10 forecast, ORBIT achieves a 1.26\% better wACC than ClimaX. 

It is important to note that the FourCastNet was trained and fine-tuned on ERA5 data and used a higher spatial resolution of 0.25 degree. ClimaX was fine-tuned on each atmospheric variable separately with specifically tailored hyperparameters. In contrast, ORBIT was pre-trained on CMIP6 model simulation data and fine-tuned on ERA5 data at a lower resolution of 1.40625 degree, predicting all four atmospheric variables together as a single task. Despite these differences, ORBIT still demonstrates competitive performance in 1-day forecasts and superior performance in longer-time forecasting compared to other state-of-the-art models.

\subsection{Fine-Tuning Data Efficiency}

 \begin{figure}
\centering
    \includegraphics[width=0.9\linewidth]{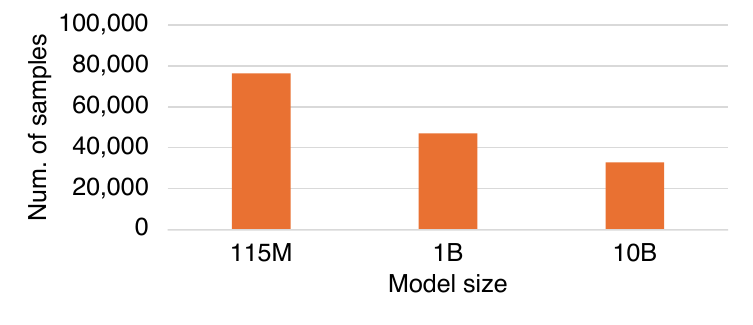}
    \caption{The number of the ERA5 samples processed by ORBIT fine-tuning models with different sizes until the test score converged for a 30-day prediction. As the model size increases, the number of samples needed to process is decreasing, indicating improved data efficiency with larger models.}
    \label{fig:model-samples}
\end{figure}

In Fig.~\ref{fig:model-samples}, we investigate fine-tuning data efficiency in relation to model sizes. We count the number of samples processed by the ORBIT models of three different sizes (113 million, 1 billion, and 10 billion parameters) until the test score converged to similar values. We use wACC as the test measurement and count the number of the ERA samples needed to converge during 30-day fine-tuning task. Our results demonstrate that as model size increases, the number of samples needed decreases, indicating greater fine-tuning data efficiency with larger models. The 115 million parameter model required about 76,000 samples to converge, while the 1 billion parameter model needed around 47,000 samples. Remarkably, the 10 billion parameter model achieved convergence with just 32,800 samples. This represents a significant reduction in the number of samples required, by approximately 38\% for the 1 billion parameter model and 57\% for the 10 billion parameter model compared to the smallest 115 million parameter model. 
These findings highlight a critical advantage of using larger models in terms of fine-tuning data efficiency. The reduction in the number of samples required for convergence as model size increases suggests that larger models are more effective in learning from smaller amounts of data. This efficiency can potentially lead to significant cost and time savings in fine-tuning processes, making it a valuable consideration for deploying large-scale AI models in various down stream climate and weather applications.

\section{Implications}
\label{sec:implications}

Earth system predictability confronts challenges due to the complex nature of environmental systems and the myriad of influencing variables. This complexity demands robust, adaptable, and highly scalable computational models to enhance our predictive capabilities and deepen our understanding of the Earth system processes. To that end, AI foundation models for climate are emerging as a promising approach in recent years. The contributions of the ORBIT model significantly impact both the field of AI for climate science and the future of HPC systems and applications.

Notably, the ORBIT's exaFLOP computing throughput performance marks a significant achievement for AI-based Earth system modeling. This capability provides the foundation for a rapid computing framework that supports ORBIT's large model size, extensive number of channels, and large number of training datasets.
With the ORBIT model’s capacity to use large ViT architecture with up to 113 billion parameters, incorporate 91 variables/channels, and pre-train on 10 model datasets from the CMIP6 project, a substantial advancement in AI-based Earth system modeling has been achieved. The model's ability to process large ViT model and a wide variety of input channels allows it to capture more complex interactions and feedback mechanisms within the Earth's climate system. This enhanced model complexity is crucial for improving the accuracy of predictions related to weather forecasting, sub-seasonal to seasonal predictions, and long-term climate projections. As climate change continues to affect global ecosystems and human societies, precise predictions become increasingly vital for planning and adaptation strategies.

The ORBIT model’s effective scaling up to 113 billion parameters on a non-NVIDIA platform, such as the Frontier supercomputer, and with up to 1.6 exaFLOP sustained computing throughput also marks a considerable advance in the use of HPC resources for complex AI tasks. With up to 113 billion parameters, ORBIT represents the largest dense vision transformer to date, surpassing the existing largest dense vision transformer by five-fold~\cite{dehghani2023scaling}, and a thousandfold larger than the largest foundational climate model, ClimaX~\cite{nguyen2023climax}, and two hundred times larger than task-specific Stormer model~\cite{stormer}.
This success demonstrates that the innovation in scaling algorithms can overcome challenges such as less advanced software stacks and limited interconnect capabilities. It also promotes a more inclusive approach to HPC system development that does not depend exclusively on one type of hardware/software stack, thereby enabling diversity in hardware utilization and encouraging the development of more adaptable and robust HPC systems.

Furthermore, the introduction of Hybrid Sharded Tensor-Data Orthogonal Parallelism in ORBIT showcases the deeper integration of AI and HPC. This approach optimizes the distribution of model parameters across multiple GPUs, boosting the efficiency and scalability of AI models and achieving 1.6 exaFLOPs sustained computing throughput. Such progress is essential for the future of AI applications in science and industry, where the demand for processing large-scale, complex datasets continues to grow. Efficiently scaling AI models on diverse HPC platforms will likely drive further innovations and a broader adoption of AI technologies across various domains. Importantly, this approach is not reliant on specialized HPC architectures, making it applicable on a broad range of HPC systems.

Besides climate modeling, ORBIT provides a template for other scientific fields that process large image datasets. Fields such as astrophysics, material science, biology, and complex computational fluid dynamics simulation could benefit from applying the scaling techniques developed for ORBIT. This methodological exchange enhances the overall capability of HPC systems to address a broader array of scientific questions, extending the frontiers of AI and HPC research.

In a nutshell, the achievements of the ORBIT model has the potential to advance the field of Earth system predictability with improved methodological AI and HPC innovations and broaden the capabilities of AI and HPC for climate modeling. These advances promote a more cohesive approach to addressing some of the most urgent computational and environmental challenges of our era.

\section*{Acknowledgments}
The authors thank Verónica G. Melesse Vergara, Mallikarjun (Arjun) Shankar and Bronson Messer for their support of high performance computing resources. Additionally, we thank Vishwas Rao for his valuable feedback to the development of this paper. 
This research used resources of the Oak Ridge Leadership Computing Facility at the Oak Ridge National Laboratory (ORNL), which is supported by the Office of Science of the U.S. Department of Energy (DOE).
This research was primary supported by the ORNL's AI Initiative sponsored by the Director's Research and Development Program at ORNL, additionally supported by the BER-ASCR SciDAC Program in the DOE, and by Dan Lu's DOE Early Career Project sponsored by the DOE BER program. 
This paper describes objective technical results and analysis. Any subjective views or opinions that might be expressed in the paper do not necessarily represent the views of the U.S. Department of Energy or the United States Government.

\bibliographystyle{IEEEtran}
\bibliography{main}

\end{document}